\documentclass[12pt,twoside]{article}
\usepackage{mathpazo}
\usepackage[mathscr]{eucal}
\usepackage{amsmath,amsfonts,amssymb,amsthm,mathabx}
\usepackage[matrix,arrow]{xy}
\usepackage{times}
\usepackage{pdfsync}
\usepackage{graphicx}
\usepackage{epstopdf}
\usepackage{dcolumn}
\usepackage{hyperref}
\usepackage{color}

\newcommand{\Vect}{\mathrm{Vect}\left(S^1\right)}

\def\be{\begin{eqnarray}}
\def\ee{\end{eqnarray}}
\def\beann{\begin{eqnarray*}}
\def\eeann{\end{eqnarray*}}
\def\beq{\begin{equation}}
\def\eeq{\end{equation}}
\def\ba{\begin{array}}
\def\ea{\end{array}}
\def\ben{\begin{enumerate}}
\def\een{\end{enumerate}}
\def\bea{\begin{eqnarray}}
\def\eea{\end{eqnarray}}

\def\5{\bar }
\def\6{\partial }
\def\7{\hat }
\def\4{\tilde }

\def\d_Vphi{\text{d}_V\hspace{-0.06em}\phi}
\def\d_Vphibar{\text{d}_V\hspace{-0.06em}\bar\phi}
\def\d_Vxi{\text{d}_V\hspace{-0.06em}\xi}

\advance\textheight\topskip
\renewcommand{\tilde}{\widetilde}
\renewcommand{\hat}{\widehat}

\renewcommand{\simeq}{\cong}

\newcommand{\dd}{\partial}
\renewcommand{\d}{\partial}

\newcommand{\binner}[2]{%
  {\langle}\kern-4.15pt{\langle}#1{,}\,#2{\rangle}\kern-4.15pt{\rangle}}

\newcommand{\half}{\frac{1}{2}}

\newcommand{\ffrac}[2]{\raisebox{.5pt}%
  {\footnotesize$\displaystyle\frac{#1}{#2}$}\kern1pt}

\newcommand{\dover}[2]{\ffrac{\dd #1}{\dd #2}}

\newcommand{\ddl}[2]{\ffrac{\dd #1}{\dd #2}}

\newcommand{\RR}{\mathbb{R}}

\def\cA{\mathcal{A}}
\def\cB{\mathcal{B}}

\def\cF{\mathcal{F}}

\def\cJ{\mathcal{J}}

\def\cM{\mathcal{M}}

\def\cP{\mathcal{P}}

\numberwithin{equation}{section} \makeatletter
\@addtoreset{equation}{section}

\begin{document}

\author{G.~Barnich, G.~Giribet, M.~Leston}

\title{Chern-Simons action for inhomogeneous Virasoro group
  as extension of three dimensional flat gravity}

\date{}

\def\mytitle{Chern-Simons action for inhomogeneous Virasoro group
  as an extension of three dimensional flat gravity}

\pagestyle{myheadings} \markboth{\textsc{\small Barnich,
 Giribet, Leston}}{%
\textsc{\small Chern-Simons action, Virasoro algebra \& extended 3d 
    flat gravity}} \addtolength{\headsep}{4pt}


\begin{centering}

  \vspace{1cm}

  \textbf{\Large{\mytitle}}

\vspace{1.5cm}

  {\large Glenn Barnich$^{a}$, Gast\'on Giribet$^{a,b,c}$ and Mauricio
    Leston$^{d}$}

\vspace{1.5cm}

\begin{minipage}{.9\textwidth}\small \it \begin{center}
    $^a$ Physique Th\'eorique et Math\'ematique, Universit\'e Libre de
    Bruxelles and International Solvay Institutes \\ Campus Plaine
    C.P. 231, B-1050 Bruxelles, Belgium. \end{center}
\end{minipage}

\vspace{.5cm}

\begin{minipage}{.9\textwidth}\small \it \begin{center}
    $^b$ Universidad de Buenos Aires FCEN-UBA and IFIBA-CONICET,
    Ciudad Universitaria, Pabell\'on I, 1428, Buenos Aires,
    Argentina. \end{center}
\end{minipage}

\vspace{.5cm}

\begin{minipage}{.9\textwidth}\small \it \begin{center}
    $^c$ Instituto de F\'{\i}sica, Pontificia Universidad Cat\'{o}lica
    de Valpara\'{\i}so, 
    Casilla 4059, Valpara\'{\i}so, Chile.\end{center}
\end{minipage}

\vspace{.5cm}

\begin{minipage}{.9\textwidth}\small \it \begin{center}
    $^d$ Instituto de Astronom\'{\i}a y F\'{\i}sica del Espacio
    IAFE-CONICET, Ciudad Universitaria, Pabell\'on IAFE, 1428 C.C. 67 
    Suc. 28, Buenos Aires, Argentina. \end{center}
\end{minipage}

\vspace{.5cm}

\end{centering}

\vspace{1cm}

\begin{center}
  \begin{minipage}{.9\textwidth}
    \textsc{Abstract}. We initiate the study of a
    Chern-Simons action associated to the semi-direct sum of the
    Virasoro algebra with its coadjoint representation. This model
    extends the standard Chern-Simons formulation of three dimensional
    flat gravity and is similar to the higher-spin extension of three
    dimensional anti-de Sitter or flat gravity. The extension can also
    be constructed for the exotic but not for the cosmological
    constant deformation of flat gravity.
  \end{minipage}
\end{center}

\vfill

\thispagestyle{empty}
\newpage

\tableofcontents

\section{Introduction}
\label{sec:introduction-1}

Higher spin extensions of ${\rm AdS}_3$ gravity \cite{Blencowe:1988gj}
have attracted a lot of attention recently (see
e.g.~\cite{Henneaux:2010xg,Campoleoni:2010zq,%
  Gaberdiel:2010ar,Gaberdiel:2010pz,Gaberdiel:2012uj}) as they allow
one to probe novel aspects of the ${\rm AdS/CFT}$ correspondence
beyond the purely gravitational sector in the most tractable setting
of three bulk dimensions. More recently, such higher spin extensions
have also been constructed for three dimensional flat gravity
\cite{Afshar2013,Gonzalez:2013oaa}.

A key feature of the Chern-Simons formulation of pure gravity in flat
space \cite{Achucarro:1987vz,Witten:1988hc} is the fact that the
action does not involve the Killing metric like it does for
semi-simple Lie algebras, but the non-degenerate invariant pairing
that exists between the translation and the rotation generators in
three dimensions. More generally, such a metric, and thus also a
Chern-Simons model, exists for all algebras
$\mathfrak{g}\ltimes_{\rm ad^*}\mathfrak{g^*}$ that are the
semi-direct sum of a given algebra $\mathfrak{g}$ with its coadjoint
representation $\mathfrak{g^*}$ embedded as an abelian ideal
\cite{Ashtekar:1989qc,Romano:1991up}.

Since the Lorentz algebra $\mathfrak{so}(2,1)$ is isomorphic to
$\mathfrak{sl}(2,\mathbb R)$, probably the most obvious
infinite-dimensional extension of flat gravity consists in replacing
$\mathfrak{sl}(2,\mathbb{R})$ by the Virasoro algebra
$\mathfrak{vir}$. Such a Chern-Simons Virasoro model involves all
ingredients that are used when applying the orbit method to the study
of representations of the Virasoro algebra (see
e.g.~\cite{Witten:1987ty,Bakas:1988mq,Taylor:1992xt,Balog:1997zz} for
considerations in the physics literature). One might hope that the
model can be useful in this context, in the same way as the Poisson
sigma model \cite{Ikeda:1993fh,Schaller:1994es} is relevant to the
problem of quantizing Poisson manifolds
\cite{Kontsevich,Cattaneo:1999fm}.

A further motivation to study the model comes from the representation
theory of the symmetry algebra of asymptotically flat three
dimensional spacetimes \cite{Ashtekar1997} (see also
\cite{Barnich:2006avcorr}). This algebra is given by the BMS algebra
in three dimensions, which is the semi-direct sum of the algebra of
vector fields on the circle with its adjoint representation embedded
as an abelian ideal,
$\mathfrak{bms}_3=\Vect\ltimes_{\rm ad}\Vect_{\rm ab}$. Because of
theorems available on unitary irreducible representations of finite
dimensional Lie algebras and groups with this particular structure
(see e.g~\cite{A.O.Barut702}), it is natural to study induced
representations, for which it is again the coadjoint representation of
the Virasoro algebra with its orbits and little groups that is
relevant \cite{Barnich:2014kra}. Finally, the solution space of
three-dimensional asymptotically anti-de Sitter spacetimes is also
classified by (two copies of) the coadjoint orbits of the Virasoro
group (see e.g.~\cite{Garbarz:2014kaa,Barnich:2014zoa} for recent
discussions). In the flat case, the coadjoint orbits of the
${\rm BMS}_3$ group are controlled by Virasoro coadjoint orbits in a
standard way appropriate to semi-direct products
\cite{Barnich:2015uva}.

After reviewing the formulation of Chern-Simons theories for
inhomogeneous groups, we start our analysis of the model with a
detailed discussion of how flat gravity and its solutions are included
in this infinite-dimensional extension. We will show that gravity
solutions can be included in the extended model at the price of
turning on additional non-gravity modes that, however, do not
backreact on the geometry.

We then study its consistent deformations. This is most conveniently
done in the context of the Batalin-Vilkovisky antifield formalism
~\cite{Batalin:1981jr,Batalin:1983wj,%
  Batalin:1983jr,Batalin:1984ss,Batalin:1985qj} (see also
\cite{Henneaux:1992ig,Gomis:1995he} for reviews) where they are
controlled by the cohomology of the BV differential in the space of
local functionals in ghost number $0$ \cite{Barnich:1993vg} (see also
\cite{Barnich:2000zw}). In particular, in the case of
$\mathfrak{iso(2,1)}$, the deformations provided in
\cite{Witten:1988hc} have been studied from this point of view in
\cite{Barkallil:2002fp}, section 7. More generally, in any ghost
number, the BV cohomology in the space of local functionals is locally
isomorphic to the Chevalley-Eilenberg cohomology of the Lie algebra
with which the Chern-Simons theory is constructed. This is a
particular case of a general result valid for AKSZ sigma models
\cite{Alexandrov:1997kv} for which the cohomology of the BV
differential in the space of local functionals is locally isomorphic
to the cohomology of the target space differential
\cite{Barnich:2009jy}. An interesting feature is that the other
primitive generator in degree $3$ of the cohomology ring of $\Vect$,
appears in this deformation.

We end with considerations on the dual boundary theory. Chern-Simons
theories on a solid cylinder induce chiral Wess-Zumino-Witten theories
on their boundary \cite{Witten:1988hf,Elitzur:1989nr}. For the
Chern-Simons formulation of three dimensional gravity, the boundary
theory has been studied in the flat case in \cite{Salomonson:1989fw},
in the anti-de Sitter case with non trivial asymptotics in
\cite{Coussaert:1995zp}, and in the flat case with non trivial
asymptotics at null infinity in \cite{Barnich:2013yka}. For the
Chern-Simons Virasoro model, one ends up with a chiral Wess-Zumino
model with a current algebra determined by
$\mathfrak{vir}\ltimes_{\rm ad^*}\mathfrak{vir^*}$. A classical
Sugawara construction then manifestly shows that the model is
conformally invariant.

\section{Chern-Simons model for inhomogeneous groups}
\label{sec:introduction}

The Chern-Simons formulation of three dimensional flat gravity is
based on the Poincar\'e algebra $\mathfrak{iso}(2,1)$. If the
generators are denoted by $P_a,J_b$, the invariant inner product
required to construct the action is given by $<P_a,J_b>=\eta_{ab}$,
$<P_a,P_b>=0=<J_a,J_b>$ with $\eta_{ab}$ the Minkowski metric in three
dimensions. This construction can be generalized by replacing
$\mathfrak{so}(2,1)$ with a generic Lie algebra $\mathfrak g$ : the
total Lie algebra is the semi-direct sum of $\mathfrak g$, whose
generators are denoted by $e_a$, with its coadjoint representation
embedded as an abelian ideal,
\begin{equation}
  [e_a,e_b]=f^c_{ab} e_c,\quad [e_a,e^{*b}]=-f^b_{ac}e^{*c},\quad [e^{*a}e^{*b}]=0.
\end{equation}
The reason why it is always possible to construct a Chern-Simons
action for such a Lie algebra is the existence of the non-degenerate
invariant inner product
\begin{equation}
  \label{eq:1}
  <e_a,e^{*b}>=\delta^b_a,\quad <e_a,e_b>=0=<e^{*a},e^{*b}>.
\end{equation}

\section{Chern-Simons Virasoro model}
\label{sec:chern-simons-viras}

What we want to do here is use as Lie algebra $\mathfrak g$ the
Virasoro algebra $\mathfrak{vir}$ with generators $L_m,Z$,
\begin{equation}
  \label{eq:2}
  [L_m,L_n]=(m-n)L_{m+n}+\frac{Z}{12}\delta_{m+n}^0m(m^2-1),\quad [L_m,Z]=0,
\end{equation}
so that
\begin{equation}
\begin{split}
  \label{eq:3}
&  [L_m,L^{*n}]=(n-2m)L^{*n-m},\quad
  [L_m,Z^*]=-\frac{1}{12}\, m(m^2-1)L^{*-m},\\ & [Z,L^{*n}]=0=[Z,Z^*],\quad
  [L^{*m},L^{*n}]=0=[L^{*m},Z^*].
\end{split}
\end{equation}
If the associated gauge field is denoted by $A=A^A_\mu\, T_A\,
dx^\mu$, with $x^\mu$ local coordinates on a three-dimensional
manifold $\cM_3$ and $T_A=(L_m,Z,L^{*m},Z^*)$, the Chern-Simons action is
\begin{equation}
  \label{eq:4}
  S_0[A]=\kappa \int_{\cM_3} \half < A,dA +\frac{2}{3} A^2>.
\end{equation}
In more details, let $A=A^mL_m+D Z+B_m L^{*m}+CZ^*$. The associated
curvatures are
\begin{equation}
\begin{split}
& F^m=dA^m+\half f^m_{nk} A^nA^k,\quad F^Z=dD+\frac{1}{24}m(m^2-1)A^{-m}
A^m,\\ & F_m=dB_m-f^k_{nm}A^n
B_k+\frac{1}{12}m(m^2-1)A^{-m}C, \quad F_{Z^*}=dC,\label{eq:14}
\end{split}
\end{equation}
where $f^k_{mn}=(m-n)\delta^k_{m+n}$. The equations of motions are
equivalent to requiring these curvatures to vanish.

This theory can also be understood as a three dimensional
BF-type theory for the Virasoro algebra. Indeed, through integrations
by parts, action \eqref{eq:4} can be rewritten as
\begin{equation}
  S_0=\kappa\int_{\cM_3} (B_m F^m+ C F^Z). \label{eq:6}
\end{equation}

Finally, introducing an additional circle $S^1$ with coordinate
$\phi\in [0,2\pi)$, elements of $\mathfrak{vir}$ are pairs
$(v,-i a)$ with $v=f(\phi)\d_\phi$ vector fields on the circle and
$a$ a real number. In particular, $L_m=(ie^{im\phi}\d_\phi, 0)$
for $m\neq 0$, $L_0=(i\d_\phi,\frac{1}{24})$, $Z=(0,1)$. The
commutation relations are given by
\begin{equation}
  \label{eq:18}
  [(v_1,-i
  a_1),(v_2,-ia_2)]=\Big((f_1f_2'-f_2f_1')\d_\phi,-\frac{i}{48\pi}\int^{2\pi}_0
  d\phi ( f'_1 f_2''-f'_2 f_1'' )\Big).
\end{equation}
Coadjoint vectors are given by pairs $(u,it)$, where
$u=h(\phi)d\phi^2$ is a quadratic differential and $t$ a real
number. In these terms, the invariant inner
  product is given by
\begin{equation}
\langle (u,it),(v,-ia)\rangle= \int^{2\pi}_0d\phi\, h f +at,\label{eq:19}
\end{equation}
with coadjoint action
\begin{equation}
ad^*_{(f,-ia)} (u,it)=\big((f h'+2 f' h
-\frac{t}{24\pi} f''')d\phi^2,0)\label{eq:20}.
\end{equation}
In particular, $L^{*m}=(-\frac{i e^{-im\phi}}{2\pi},0)$,
$Z^*=(\frac{i}{48\pi},1)$.

Let $(\cA,-iD)$ and $(\cB,iC)$ be one forms on $\cM_3$ with values in
$\mathfrak{vir}$ respectively $\mathfrak{vir}^*$ and let
$(\cF,-i F^Z)=d(\cA,-iD)+\half[(\cA,-iD),(\cA,-iD)]$. Instead of an
infinite sum over $m$, the Chern-Simons
action \eqref{eq:6} can then be written with an additional integral
over $\phi$ as 
\begin{equation}
  \label{eq:21}
  S_0=\kappa\int_{\cM_3} \langle (\cB,iC),(\cF,-iF^Z)\rangle,
\end{equation}
and interpreted as a field theory on $\cM_3\times S^1$. 

Interestingly, the construction of Chern-Simons theories for
inhomogeneous groups as described in section
\ref{sec:chern-simons-viras} can also be done by using as
$\mathfrak g$ the semi-direct sum of the Virasoro algebra with the
affine Kac-Moody algebra for $\mathfrak{sl}(2,\mathbb R)$. A
Chern-Simons theory of precisely this type (but without the Virasoro
central extension) has appeared previously in the context of a
Kaluza-Klein reduction for four dimensional gravity on a circle in
\cite{Hohm2006,Hohm2006a}.

\section{Extension of flat gravity}
\label{sec:extens-flat-grav}

The Virasoro Chern-Simons model may be interpreted as describing the
non trivial coupling of an infinite number of additional gauge fields
to gravity in three dimensions since the action reduces to the
Einstein-Hilbert one by putting to zero the gauge fields associated with
the generators $Z,Z^*$ and $L_m,L^{*m}$ for $m\neq -1,0,1$. Indeed,
when all gauge fields besides $A^m, B_m, m,= -1,0,1$ are switched off,
we get
\begin{equation}
  S_0=\kappa\int B_m F^m,\quad m,n,k\in
  -1,0,1. \label{eq:6a}
\end{equation}
The standard gravitational formulation is recovered through a change
of basis: $A^m L_m+B_m L^{*m}=\omega^a J_a +e^a P_a$, with
$a,b= 0,1,2$, $J^a=-\half \epsilon^{abc} J_{bc}$,
$[J_a,J_b]=\epsilon_{abc} J^c$, $[J_a,P_b]=\epsilon_{abc}P^c$,
$[P_a,P_b]=0$, $\epsilon_{012}=1$, where $\omega ^a$ and $e^a$ are the
spin connection and vielbein 1-forms, respectively.  Indices are
lowered and raised with $\eta_{ab}={\rm diag}(-1,1,1)$ and its inverse
while
\begin{equation}
\begin{split}
& l_{-1}=J_0-J_1,\quad
l_1=J_0+J_1,\quad l_0=-J_2,\\ 
& P_0=-l^{*1}-l^{*-1},\quad P_1=
l^{*1}-l^{*-1},\quad P_2=-l^{*0}. \label{eq:22}
\end{split}
\end{equation}
and action \eqref{eq:6a} coincides with the Einstein-Hilbert
action in terms of dreibeins and spin connections,
\begin{equation}
  \label{eq:13}
   S_0[e,\omega]=\frac{1}{16\pi G}\int e e_a^\mu e_b^\nu
   R^{ab}_{\mu\nu} \,d^3 x,
\end{equation}
provided $\kappa=-\frac{1}{8\pi G}$.  

The question is then whether any solution to the gravity equations of
motions can be lifted to a solution of the extended system. More
generally, consider a Chern-Simons theory based on
$\mathfrak{g}\ltimes_{\rm ad^*}{\mathfrak g}^*$ and let $\mathfrak
h\subset \mathfrak g$ be a subalgebra. In the gravity case, $\mathfrak
h=\mathfrak{sl}(2,\RR)$, while $\mathfrak g=\mathfrak{vir}$. Can any
solution of the Chern-Simons theory based on $\mathfrak{h}\ltimes_{\rm
  ad^*}{\mathfrak h}^*$ be lifted to a solution of the
Chern-Simons theory based on $\mathfrak{g}\ltimes_{\rm ad^*}{\mathfrak
  g}^*$ ?

Answering this question is not completely trivial because
$\mathfrak{h}\ltimes_{\rm ad^*}{\mathfrak h}^*$ is not a subalgebra of
$\mathfrak{g}\ltimes_{\rm ad^*}{\mathfrak g}^*$. Indeed, let us denote
by $e_\gamma$ generators of $\mathfrak h$ and by
$e_\Gamma=(e_\gamma,e_C)$ generators of $\mathfrak g$. The equations
of motion of the theory based on $\mathfrak g$ are
\begin{equation}
  \label{eq:23}
  dA^\Gamma+\half f^\Gamma_{\Delta\Sigma} A^\Delta A^\Sigma=0,\quad 
dB_\Gamma-f^\Delta_{\Sigma\Gamma} A^\Sigma B_{\Delta}=0, 
\end{equation}
while those of the theory based on $\mathfrak h$ are similar but with
upper case Greek indices replaced by lower case ones. Now, while it is
true that given a solution of the theory based on $\mathfrak h$, the
first set of equations in \eqref{eq:23} can always be solved by
setting to zero the all gauge fields $A^C$ since
$f^{S}_{\gamma\delta}=0$, this cannot be done for the second set of
equations since switching off all fields $B_C$ results in the
additional constraints $f^{\gamma}_{\delta C}A^\delta B_\gamma=0$ on
the fields of the theory based on $\mathfrak h$.

Let us then provide a formal argument showing that every solution of
the theory based on $\mathfrak h$ can be lifted to a solution of the
theory based on $\mathfrak g$ by keeping the fields $B_\gamma$
unchanged and suitably turning on additional fields $B_C$. Indeed,
locally, the general solution of the theory based on $\mathfrak h$ is
given by
\begin{equation}
A=h^{-1}dh,\quad 
B=h^{-1}(d C_\gamma e^{*\gamma}) h,\label{eq:24}
\end{equation}
with $h$ a group element associated to $\mathfrak h$ and $C_\gamma(x)$
arbitrary spacetime-dependent functions. The lift of the solution is
then simply obtained by considering the formal group element
$h=\text{e}^{k^\alpha e_\alpha}$ with spacetime dependent functions
$k^\alpha(x)$ to be an element of the group associated to $\mathfrak
g$. This gives
\begin{equation}
B=dC_\gamma \text{e}^{-[k^\alpha e_\alpha,\cdot]}e^{*\gamma}=
dC_\gamma (e^{*\gamma}+k^\alpha f_{\alpha\Delta}^\gamma
e^{*\Delta}+\half k^\alpha k^\beta f^\gamma_{\beta\Sigma}
f^{\Sigma}_{\alpha\Delta} e^{*\Delta}+\dots )\label{eq:25}. 
\end{equation}
The reason why $B_\gamma$ is unchanged is that 
\begin{equation*}
[\mathfrak
h,[\mathfrak h,\dots[\mathfrak h,\mathfrak
g^*]\dots]]|_{h^*}=[\mathfrak h,[\mathfrak h,\dots[\mathfrak
h,\mathfrak h^*]\dots]]_{{\mathfrak{h}\ltimes_{\rm
      ad^*}\mathfrak{h^*}}}
\end{equation*}
since $\mathfrak h$ is a subalgebra of
$\mathfrak g$.

Note however that the exponential map does not define a local chart
close to the identity for the diffeomorphism of the circle or the
Virasoro group. The above argument then only makes sense if one thinks
about these groups formally, as done in the current context for
instance in \cite{Aldaya:1989ra,Taylor:1992xt}.

More concretely, let us consider for instance the general solution of
three-dimensional flat gravity in BMS gauge \cite{Barnich:2010eb}. The
associated vielbein and spin connections are given in
\cite{Barnich:2013yka}. Explicitly, when taking into account
\eqref{eq:22}, they translate to
\begin{equation}
\begin{split}
A^{\pm1}=(\frac{M}{2}\mp 1)d\phi,\quad
A^{0}=0,\\
B_{\pm1}=-\frac{1}{2}((\frac{M}{2}\pm 1)du-dr + Nd\phi),\quad
B_0=-rd\phi. \label{BMSsol}
\end{split}
\end{equation}
Here $M$ and $N$ are defined in terms of two arbitrary function of
$\phi,$ $\Theta(\phi)$ and $\Xi(\phi)$ through $M=\Theta$, $N=\Xi +
\frac{u}{2}\Theta'$.  

Acccording to the previous considerations, we can assume $A^m=0=D$ for
$\mid{m}\mid>1$. Let us then show that it is enough to turn on in
addition $B_{\pm 2}$ in order to extend the above solutions.

Indeed, if in addition $B_m=0=C$ for $\mid{m}\mid>2$, the equations
for $dB_m,dC$ are automatically fulfilled for $\mid{m}\mid>3$. The
equations for $\mid{m}\mid=2,3$ read explicitly
\begin{equation}
\begin{split}
dB_2 +3 A^{-1}B_{1} + A^1B_3 +2A^0B_2=0,\\
dB_{-2} - 3 A^{1}B_{-1} -A^{-1}B_{-3} - 2A^0B_{-2}=0,\\
A^{-1}B_{2} =0,\quad A^{1}B_{-2}=0.
\end{split} 
\end{equation}
Choosing $B_2= f_{+}A^{-1}$ and $B_{-2}=f_{-}A^{1}$ for some functions
$f_+$ and $f_{-}$ allows one to satisfy the last 2 equations, while
the first 2 turn into a system of partial differential equations for
$f_+$ and $f_-$,
\begin{eqnarray}
d(f_+A^{-1}) +3 A^{-1}B_{1} + 2f_+A^0A^1=0, \label{eqfplus}\\
d(f_{-}A^{1}) - 3 A^{1}B_{-1}- 2f_{-}A^0A^{-1}=0.\label{eqfminus}
\end{eqnarray}
By using that for \eqref{BMSsol},
$A_0=0,dA^{\pm{1}}=0$, they reduce to 
$df_{\pm}=\mp3B_{\pm1}$. The associated integrability conditions 
$dB^{\pm{1}}=0$ are satisfied for \eqref{BMSsol} and 
\begin{equation}
f_{\pm}=\pm\frac{3}{2}(\frac{u}{2}\Theta \pm u + \tilde{\Xi}-r),
\end{equation}
where $\tilde{\Xi}'=\Xi$.

Then, we see explicitly that the field equations for the gravity modes
are indeed satisfied. The gravity modes, even when they source the
extra modes with $|m|>1$, do it in such a way that the latter, if
originally switched off, do not backreact on Einstein's field
equations.

\section{Deformations of the Chern-Simons Virasoro model}
\label{sec:gener-deform-viras}

Studying consistent deformations of the Chern-Simons action
\eqref{eq:4} in the context of the Batalin-Vilkovisky antifield
formalism involves several steps.

(i) All infinitesimal deformations must belong to the cohomology group
$H^{0,3}(s|d)$, where $s$ is the antifield dependent BRST differential
associated with the Batalin-Vilkovisky master action for the Virasoro
Chern-Simons action \eqref{eq:4}.

(ii) For Chern-Simons theories, $H^{0,3}(s|d)$ is isomorphic to the
Chevalley-Eilenberg cohomology of the associated Lie algebra.

For simplicity, let us first restrict ourselves to the non-centrally
extended case of the Lie algebra $\mathfrak{witt}\ltimes
\mathfrak{witt}^*$ with generators $l_m, l^{*n}$, so that $c,Z,Z^*=0$.
In this case, $H^{0,3}(s|d)\simeq H^3(\mathfrak{witt}\ltimes
\mathfrak{witt}^*)$.

(iii) The Hochschild-Serre analysis \cite{Hochschild1953} can be
  adapted to the current problem. Denoting the generators of
  {$\wedge$}$(\mathfrak{witt}\ltimes \mathfrak{witt}^*)^*$ by $\eta^m$ and
  $C_{m}$, the Chevalley-Eilenberg coboundary operator can be written
  as
  \begin{equation}
    \label{eq:9}
    \gamma = -\half \eta^m\eta^nf^k_{mn}\dover{}{\eta^k}+\eta^m\rho(l_m),\quad
    \rho(l_m)=f^k_{mn}C_{k}\dover{}{C_{n}}.
  \end{equation}
  Let $N_\eta=\eta^{m}\dover{}{\eta^{m}}$ and $N_C=C_{m}\dover{}{C_{m}}$. Since
  $[\gamma,N_C]=0$, the problem decomposes into separate cohomology
  problems with eigenvalues $(3,0)$, $(2,1)$, $(1,2)$, $(0,3)$ for
  $(N_\eta,N_C)$. 

In degree $(3,0)$ there is but one cohomology class
  (\cite{Gelfand1968}, \cite{Fuks:1986}, Theorem 2.4.2),
  which can be written as $\beta=\frac{1}{3!}\beta_{mnk}\eta^m\eta^n\eta^k$
  with
  \begin{equation}
\beta_{mnk}=\beta_{[mnk]}=\delta^0_{m+n+k}\big(mn(m-n)
+nk(n-k)+mk(k-m)\big)\label{eq:10}.
\end{equation}
If the gauge field is denoted by $B^ml_m+A_m l^{*m}$, the BV master
action is given by
\begin{multline}
  \label{eq:11}
  S=\kappa\int [A_m dB^m +\half A_k f^k_{mn} B^mB^n+\\+\star A^{*m}(
  dC_m+f^k_{mn} B^n C_k-f^k_{mn} A_k\eta^n)+\star B^*_m( d\eta^m
  +f^m_{nk} B^n\eta^k)+\\+\half \star \eta^*_k f^k_{mn}\eta^m\eta^n-
  \star C^{*m} f^k_{mn} C_k\eta^n].
\end{multline}
We follow the conventions of \cite{Barnich:2000zw} : if
$\omega=\frac{1}{p!} dx^{\mu_1}\wedge\dots \wedge dx^{\mu_p}
\omega_{\mu_1\dots\mu_p}$,
then
$\star \omega= \frac{1}{p!(n-p)!} dx^{\mu_1}\wedge\dots \wedge
dx^{\mu_{n-p}}\epsilon_{\mu_1\dots
  mu_n}\omega^{\mu_{n-p+1}\dots\mu_n}$
where $\epsilon^{\mu_1\dots\mu_n}$ is completely skew symmetric,
$\epsilon^{01\dots n-1}=1$, and indices are lowered and raised with
$\eta_{\mu\nu}={\rm diag}(-1,1,\dots,1)$.

The associated infinitesimal deformation of the master action is
  \begin{equation}
    \label{eq:12}
    S^{(1)}=\int  \frac{1}{3!} \beta_{mnk}(B^mB^nB^k+6 {\star }
    A^{*m}B^n\eta^k+3{\star } C^{*m}\eta^n\eta^k). 
  \end{equation}
  In particular, $(S^{(1)},S^{(1)})=0$, so the infinitesimal
  deformation is by itself a complete deformation,
  $S^{\rm def}=S+\mu S^{(1)}$. It plays the role of the topological
  deformation \cite{Deser:1982vy,Deser:1981wh}, that has been called
  ``exotic'' term in \cite{Witten:1988hc}. On the level of the Lie
  algebra, it amounts to keeping the same inner product but deforming
  the commutator
  $[l_{m},l_{n}]^{\rm def}=f^k_{mn} l_k+\mu\beta_{mnk} l^{*k}$ while
  keeping all other commutation relations unchanged. That this is
  legitimate can be checked a posteriori: indeed, the inner product
  stays invariant because of the complete skew-symmetry of
  $\beta_{mnk}$, while for the Jabobi identity, only
  $[l_l,[l_m,l_n]^{\rm def}]^{\rm def}+{\rm cyclic} (l,m,n)$ is non
  trivial. Evaluating, one finds that this vanishes on account of the
  cocycle condition since
  $f^k_{mn}\beta_{klr}+f^k_{lr}\beta_{kmn}+{\rm cyclic} (l,m,n)$ is
  completely skew-symmetric in $(l,m,n,r)$.

In degree $(0,3)$, there is no class, because one would need a
skew-symmetric three index tensor that is invariant under the
coadjoint representation of the Witt algebra. The only candidate is
the unique invariant tensor $\delta_{-101}^{mnk}$ under the coadjoint
representation of $\mathfrak{sl}(2,\mathbb R)$, which is no longer
invariant under transformations generated by $l_l$ with $l\neq
-1,0,1$. Note that in the case of $\mathfrak{sl}(2,\mathbb R)$, this
is the one that is responsible for the cosmological constant
deformation. Whether there are classes in degrees $(1,2)$ and $(2,1)$
needs to be investigated.  

Let us now turn to the centrally extended case of the Lie algebra
$\mathfrak{vir}\ltimes \mathfrak{vir}^*$. 

First, note that $H^1(\mathfrak{vir})=0=H^2(\mathfrak{vir})$. Indeed,
by denoting by $C^a$ the ghosts associated to the $\mathfrak{witt}$
algebra and by $C^Z$ the ghost associated to the central element,
Chevalley-Eilenberg differential for the Witt algebra is
$\gamma=\half C^a C^b f_{ab}^c\ddl{}{C^c}$, while the one for the
Virasoro algebra is
$\gamma^T=\half C^a C^b f_{ab}^c\ddl{}{C^c}+\omega\ddl{}{C^Z}$, where
$\omega= \omega_{[ab]} C^a C^b$ is a representative of
$H^2(\mathfrak{witt})$, which is 1 dimensional.

For the Virasoro algebra, the cocyle condition in degree $1$ for an
element $k_a C^a+ k C^Z$ implies that $k\omega=-\gamma(k_a C^a)$. This
implies in turn $k=0$ and, since $H^1(\mathfrak{witt})=0$, that $k_a
C^a$ is $\gamma$ and also $\gamma^T$ exact, which gives the result.

The cocycle condition in degree $2$ splits depending on whether it
involves $C^Z$ or not.  The former piece implies that the coefficient
of $C^Z$ is a $\gamma$ cocycle in degree $1$; $H^1(\mathfrak{witt})=0$
then implies that it is a coboundary, and since there are no
coboundaries in degree $1$, that it vanishes.  The piece independent
of the $C^Z$ ghost then has to be a $\gamma^T$ and thus also a
$\gamma$ cocycle, which implies that it is given by $\omega$ up to a
$\gamma$ coboundary $\gamma \eta^1$, where $\eta^1$ depends on the
$C^a$ alone, so that $\gamma \eta^1=\gamma^T\eta_1$. But, by
construction, $\omega=\gamma^T C^Z$ which proves the result.

A similar reasoning using in addition that $\omega^2$ is a non trivial
cohomology class in $H^4(\mathfrak{witt})$ then allows one to show
that $H^3(\mathfrak{vir})$ is one dimensional and also described by
$\beta$. Indeed, for an element
$\alpha=k_{[abc]}C^aC^bC^c+k_{[ab]}C^aC^BC^Z$, the cocycle condition
$\gamma^T\alpha=0$ implies $\gamma(k_{[ab]}C^aC^B)=0$ so that
$k_{[ab]}C^aC^B=\gamma(k_a C^a)+ k\omega$. This implies that
$\alpha=k'_{[abc]}C^aC^bC^c+\gamma^T(k_a C^aC^Z)+k\omega C^Z$, and the
cocycle condition becomes $\gamma(k'_{[abc]}C^aC^bC^c)+k\omega^2=0$
which implies $k=0$ and $k'_{[abc]}C^aC^bC^c=k\beta+\gamma(l_{[ab]}C^a
C^b)=\beta+\gamma^T(l_{[ab]}C^a C^b)$. Finally,
$k\beta=\gamma^T(l'_{[ab]}C^aC^b+l_a C^aC^Z)$ implies $\gamma l_a
  C^a)=0$ which gives $l_a C^a=0$ and then $k=0$, so that $\beta$
  remains non trivial.

\section{Dual boundary theory}
\label{sec:dual-boundary-theory}

Following closely the case of $\mathfrak{iso}(2,1)$ treated in
\cite{Salomonson:1989fw,Barnich:2013yka}, the boundary theory when
$\mathfrak{so}(2,1)\simeq \mathfrak{sl}(2,\mathbb R)$ is replaced by
the Witt or Virasoro algebra is
\begin{equation}
  \label{eq:54a}
  I[\lambda,\alpha]=\frac{k}{2\pi} \int dud\phi\ \langle \dot
  \lambda\lambda^{-1},\alpha'\rangle, 
\end{equation}
where $\lambda(u,\phi)$ is a map from the cyclinder to the ${\rm
  Diff}(S^1)$/Virasoro group, $\alpha(u,\phi)$ a map whose value is an
associated coadjoint vector, and the inner product
$\langle\cdot,\cdot\rangle$ between coadjoint vectors and Lie algebra
elements replaces $2{\rm Tr}$ in \cite{Barnich:2013yka}. Note that in
that reference, the integrand contains an additional term $-{\rm
  Tr}(\lambda_{{\rm SL}(2,\mathbb R)}'\lambda_{{\rm SL}(2,\mathbb
  R)}^{-1})^2$ originating from a Gibbons-Hawking type improvement
term needed to account for the non trivial asymptotics of the spin
connection. Because of the absence of an invariant trace, this term
cannot be extended to a ${\rm Diff}(S^1)$/Virasoro group element. In
order to continue to describe an extension of gravity and include the
$u$-dependent solutions discussed at the end of section
\ref{sec:extens-flat-grav}, one should add this terms, with the group
element restricted to ${\rm SL}(2,\mathbb R)$. For simplicity, we
choose not to do so here and concentrate on the model defined by
\eqref{eq:54a}.

The equations of motion of the model are
\begin{equation}
 (\dot\lambda\lambda^{-1})'=0,\quad
D_u^{-\dot\lambda\lambda^{-1}}\alpha^\prime=0.\label{eq:44}
\end{equation}
The general solution involves a factorized group element
$\lambda=\mu(u)\nu(\phi)$, and $\alpha=\mu
(\rho(\phi)+\delta(u))\mu^{-1}$. The gauge invariance of the action is
$\lambda\to\nu(u)\lambda$, $\alpha\to\nu\alpha\nu^{-1}$, while the
global symmetries are $\lambda\to\lambda\Theta(\phi)^{-1}$,
$\alpha\to\alpha+\lambda \Sigma(\phi)\lambda^{-1}$. The infinitesimal
version of the latter are $\delta_\theta\lambda=-\lambda\theta$,
$\delta_\theta\alpha =0$, and $\delta_\sigma\lambda=0, \delta_\sigma
\alpha=\lambda \sigma 
\lambda^{-1}$  with Noether currents
\begin{equation}
  \label{eq:96}
\begin{split}
  J^0_\theta=\langle \theta,J\rangle,\quad
J=-\frac{k}{2\pi}\big[\lambda^{-1}\alpha'
\lambda\big],\quad
  J^1_\theta=0,\\
 P^0_\sigma=\langle \sigma, P\rangle,\quad P=\frac{k}{2\pi}
\lambda^{-1}\lambda',\quad
  P^1_\sigma=0.
\end{split}
\end{equation}
In terms of generators, $J_m= \langle l_m ,J\rangle$,
$\zeta=\langle Z, J\rangle$, and
$P^m=\langle l^{*m}, P\rangle$, $\zeta^*=\langle Z^*,P\rangle$, one
then reads off the current algebra form the structure constants and
the inner product, 
\begin{equation}
  \label{eq:94}
  \begin{split}
    & \{P^m(\phi),P^n(\phi')\}^*=0= \{P^m(\phi),\zeta(\phi')\}^*
    =\{P^m(\phi),\zeta^*(\phi')\}^*,\\
    &
    \{\zeta(\phi),\zeta^*(\phi')\}^*={-\frac{k}{2\pi}
    \d_\phi\delta(\phi-\phi')},\quad  
    \{\zeta(\phi),J_m(\phi')\}^*=0, \\
    &
    \{J_m(\phi),\zeta^*(\phi')\}^*=-\frac{1}{12}m(m^2-1)P^{-m}(\phi)
    \delta(\phi-\phi'),\\ 
    & \{J_m(\phi),P^n(\phi')\}^*=(n-2m)
    P^{n-m}(\phi)\delta(\phi-\phi')-\frac{k}{2\pi}
    \delta^n_m\d_\phi\delta(\phi-\phi'),\\
    & \{J_m(\phi),J_n(\phi')\}^*=\big[(m-n)
    J_{m+n}(\phi)+\frac{\zeta(\phi)}{12}
    \delta^0_{m+n}m(m^2-1)\big]\delta(\phi-\phi'). 
  \end{split}
\end{equation}
The natural combination that is available for a classical Sugawara
construction is
\begin{equation}
  \label{eq:16}
  \cP {\approx} -\frac{2\pi}{k} (J_m P^m+\zeta\zeta^*),
\end{equation}
and
\begin{equation}
  \label{eq:15}
  \{\cP(\phi),\cP(\phi')\}^*=(\cP(\phi)+\cP(\phi'))
  \d_\phi\delta(\phi-\phi'), 
\end{equation}
or, in terms of modes $\cJ_m=\int^{2\pi}_0 d\phi\ e^{im\phi} \cP$, 
\begin{equation}
  \label{eq:17}
   i\{\cJ_m,\cJ_n\}=(m-n) \cJ_{m+n},
\end{equation}
which expresses conformal invariance of the model. 

\section*{Acknowledgements}
\label{sec:acknowledgements}

\addcontentsline{toc}{section}{Acknowledgments}

This work is supported by two FNRS/MINCyT collaboration agreements.
The work of G.B.~and G.G.~is partially funded by the Fund for
Scientific Research-FNRS Belgium (convention FRFC PDR T.1025.14 and
convention IISN 4.4503.15) and by ``Communaut\'e fran\c caise de
Belgique - Actions de Recherche Concert\'ees''. The work of G.G.~is
also supported by a donation from the Solvay family.



\section*{References}
\addcontentsline{toc}{section}{References}

\renewcommand{\section}[2]{}%

\def\cprime{$'$}
\providecommand{\href}[2]{#2}\begingroup\raggedright\endgroup

\end{document}